\documentstyle[epsfig,proceedings]{crckapb} 

\begin{opening}
\title{HIGH MASS BLACK HOLES IN SOFT X-RAY TRANSIENTS}
\subtitle{Gap in Black Hole Masses ?}

\author{G.E. Brown, C.-H. Lee}
\institute{Department of Physics and Astronomy,\\
 SUNY at Stony Brook, NY 11794, USA}
\author{H.A. Bethe}
\institute{Floyd R. Newman Laboratory of Nuclear Studies,\\
        Cornell University, Ithaca, New York 14853, USA}

\end{opening}

\runningtitle{High Mass Black Holes in Soft X-Ray Transients}

\begin{document}


\newcommand{\be}{\begin{eqnarray}}
\newcommand{\ee}{\end{eqnarray}}
\def\lsim{\mathrel{\rlap{\lower4pt\hbox{\hskip1pt$\sim$}}
        \raise1pt\hbox{$<$}}} 
\def\gsim{\mathrel{\rlap{\lower4pt\hbox{\hskip1pt$\sim$}}
        \raise1pt\hbox{$>$}}} 
\def\ul{\underline}
\def\in{\indent}
\def\lin{\in\in\in}
\def\etal{{\it et al.} }
\def\eg{{\it e.g.,} }
\def\ie{{\it i.e.,} }
\def\kms{{km s$^{-1}$}}
\def\kpc{kpc}
\newcommand{\msun}{~{\rm M}_\odot}
\newcommand{\rsun}{~{\rm R}_\odot}
\long\def\beginomit#1\endomit{}

\begin{abstract}
       \footnote{
       Talk given by C.-H. Lee at the NATO Advanced Study Institute
       on ``The Neutron Star - Black Hole Connection", 
       June 7 - 18, 1999, Elounda, Crete, Greece}
We suggest that high-mass black holes; i.e., black holes of
several solar masses, can be formed in binaries with low-mass main-sequence
companions, provided that the hydrogen envelope of the massive star is
removed in common envelope evolution which begins only after the massive
star has finished He core burning. 
Our evolution scenario naturally explains the gap 
(low probability region) in the observed black hole masses.
\end{abstract}

\section{Introduction}
\label{sec0}

In this talk we suggest that high-mass black holes; i.e., black holes of
several solar masses, can be formed in binaries with low-mass main-sequence
companions, provided that the hydrogen envelope of the massive star is
removed in common envelope evolution which begins only after the massive
star has finished He core burning (Brown, Lee, \& Bethe 1999). 
That is, the massive star is in the
supergiant stage, which lasts only $\sim 10^4$ years, so effects of mass
loss by He winds are small. Since the removal of the hydrogen envelope
of the massive star occurs so late, it evolves essentially as a single
star, rather than one in a binary. Thus, we can use evolutionary
calculations of Woosley \& Weaver (1995) of single stars.

We find that high-mass black holes can be formed in the
collapse of stars with ZAMS mass $\gsim 20\msun$. 
Mass loss by winds in stars
sufficiently massive to undergo the LBV (luminous blue variable) stage
may seriously affect the evolution of stars of ZAMS $>35-40\msun$, 
we take the upper limit for the evolution of the so-called transient
sources to be $\sim 35\msun$ ZAMS mass. 
Both Portegies Zwart, Verbunt \& Ergma (1997) and
Ergma \& Van den Heuvel (1998) have suggested that roughly our
chosen range of ZAMS masses must be responsible for the transient
sources. We believe that the high-mass black hole limit of ZAMS mass 
$\sim 40 \msun$ suggested by Van den Heuvel \& Habets (1984) and
later revised to $\ge 50 \msun$ (Kaper et al. 1995) applies to
massive stars in binaries, which undergo RLOF (Roche Lobe Overflow) early
in their evolution. 

\begin{table}
   \caption[Black holes candidates with measured mass functions]{Parameters
	   of suspected black hole binaries in soft X-ray transients 
           with measured mass functions (Brown, Lee, \& Bethe 1999).
	   N means nova, XN means X-ray nova. Numbers in parenthesis indicate
	   errors in the last digits.
           }
   \label{tab1}
\begin{center}
\small
\newcommand{\ti}[1]{{\tiny #1}}
\noindent\begin{tabular}{@{}llcccc@{}}\hline
                &               &  compan. &$P_{orb}$   &$f(M_{X})$&$M_{opt}$  \\
X-ray           & other         &  type    & (d)        &  ($\msun$)    &  ($\msun$)   \\ \cline{3-6}
names           & name(s)       &  q       &$K_{opt}$   &  i          &$M_{X}$    \\
                &               &  ($M_{opt}/M_X$) & (\kms)     &  (degree)   &  ($\msun$)   \\ \hline
XN Mon 75         &\ti{V616\,Mon}&  K4 V   & 0.3230     & 2.83-2.99      & 0.53--1.22 \\
\ti{A\,0620$-$003}&\ti{N Mon 1917}& 0.057--0.077 & 443(4)     & 37--44 &  9.4--15.9 \\ 
XN Oph 77       &\ti{V2107\,Oph}&  K3 V     & 0.5213     & 4.44--4.86 &  0.3--0.6 \\
\ti{H\,1705$-$250}&             &           & 420(30)    &   60--80   &  5.2--8.6 \\ 
XN Vul 88       &\ti{QZ\,Vul}    &  K5 V   & 0.3441     &  4.89--5.13 & 0.17--0.97   \\
\ti{GS\,2000$+$251}&            & 0.030--0.054 & 520(16)    &  43--74   &  5.8--18.0 \\ 
XN Cyg 89       &\ti{V404\,Cyg}&  K0 IV     & 6.4714     & 6.02--6.12  & 0.57--0.92  \\
\ti{GS\,2023$+$338}&\ti{N Cyg 1938, 1959}& 0.055--0.065 & 208.5(7)   &   52--60    & 10.3--14.2 \\ 
XN Mus 91       &               &  K5 V     & 0.4326     & 2.86--3.16 & 0.41--1.4 \\
\ti{GS\,1124$-$683}&           &  0.09--0.17 & 406(7)     &   54--65   & 4.6--8.2 \\ 
XN Per 92       &               &  M0 V   & 0.2127(7)  & 1.15--1.27 & 0.10-0.97 \\
\ti{GRO\,J0422$+$32}&           &  0.029--0.069 & 380.6(65)  & 28--45   & 3.4--14.0 \\ 
XN Sco 94       &               &  F5-G2    & 2.6127(8)  & 2.64--2.82 & 1.8--2.5   \\
\ti{GRO\,J1655$-$40}&           &  0.33--0.37 & 227(2)     &   67--71   & 5.5--6.8 \\ 
XN               &\ti{MX 1543-475}&  A2 V     & 1.123(8)   & 0.20--0.24 & 1.3--2.6   \\
\ti{4U 1543$-$47}&                 &           & 124(4)     &  20-40    & 2.0--9.7   \\ 
XN Vel 93  &                & K6-M0   & 0.2852     &  3.05--3.29 & 0.50--0.65  \\
           &                & 0.137$\pm$ 0.015& 475.4(59) & $\sim$ 78 &  3.64--4.74  \\ 
\hline
\end{tabular}
\end{center}
\end{table}

The most copious high-mass black holes of masses $\sim 6 -7 \msun$
have been found in the transient sources such as A0620.
These have low-mass companions, predominantly of $\lsim 1 \msun$, such as
K-- or M--stars.
In the progenitor binaries the mass ratios must have
been tiny, say $q\sim 1/25$. 
Following the evolutionary scenario for the black hole
binary of De Kool et al. (1987),
we show that the reason for this small $q$-value
lies in the common envelope evolution of the binary. The smaller
the companion mass, the greater the radius $R_g$ the giant must reach
before its envelope meets the companion.
This results because 
the orbit of a low-mass companion must shrink by a large factor in
order to expel the envelope of the giant, hence the orbit must initially
have a large radius. (Its final radius must be just inside its Roche
Lobe, which sets a limit to the gravitational energy it can furnish.)
A large radius $R_g$ in turn means that the primary star must be in the
supergiant stage. Thus it will have completed its He core burning
while it is still ``clothed" with hydrogen. This prevents excessive
mass loss so that the primary retains essentially the full mass of its
He core when it goes supernova.
We believe this is why
K-- and M--star companions of high-mass black holes are favored.

\begin{figure}
\begin{minipage}[t]{0.48\textwidth}
\centerline{\epsfig{file=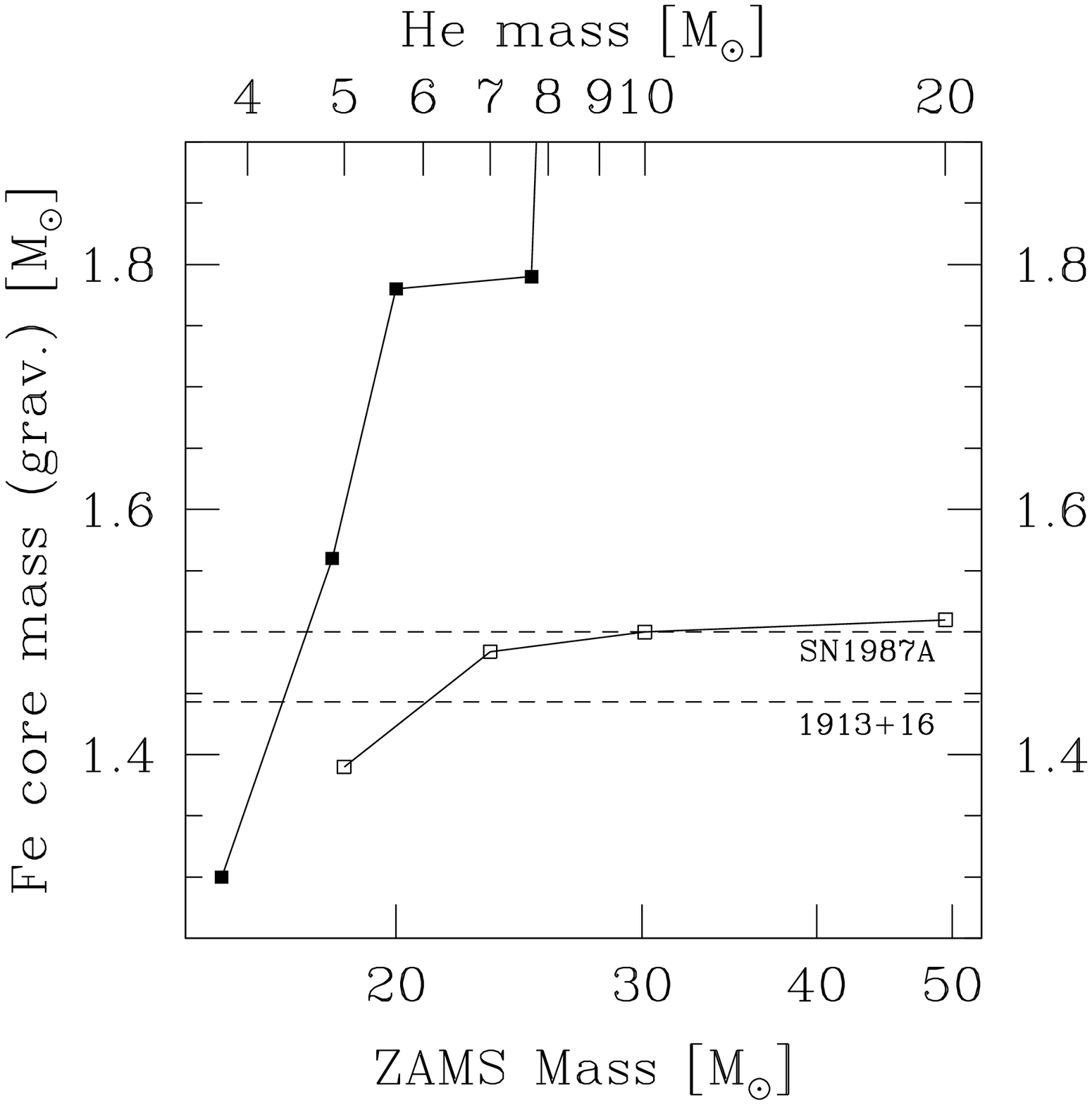,height=5cm}}
\caption{Comparison of the compact core masses resulting from the
evolution of single stars (filled symbols, Case B of Woosley \& Weaver 1995),
and naked helium stars (Woosley, Langer \& Weaver 1995) with masses equal to
the corresponding He core mass of single stars.  
}
\label{fig1}
\end{minipage}
\hfill
\begin{minipage}[t]{0.48\textwidth}
\centerline{\epsfig{file=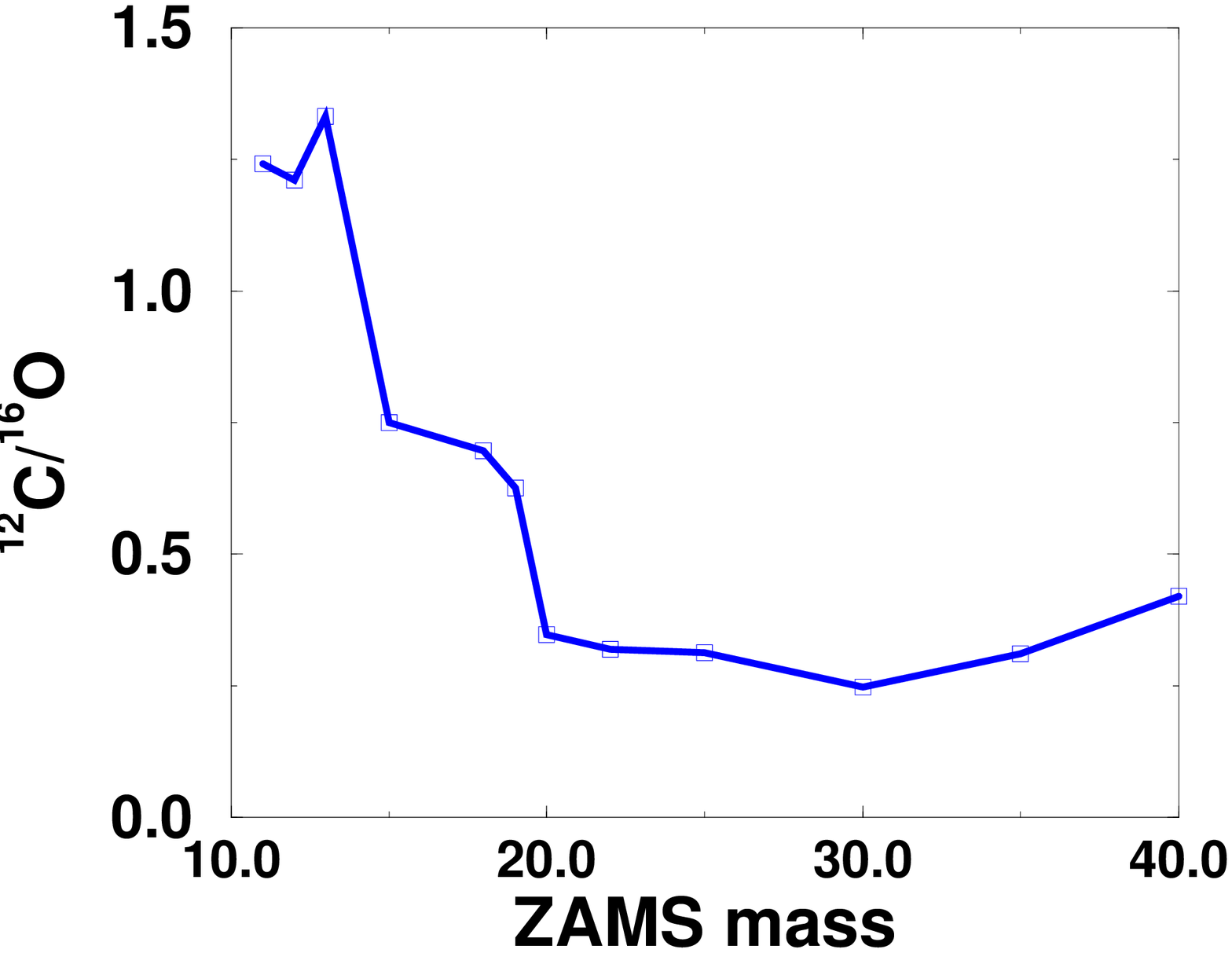,width=0.98\textwidth}}
\caption{
Ratio of production rates of $^{12}C$ and $^{16}O$
resulting from the evolution of single stars (filled symbols), case of solar
metallicity of Woosley \& Weaver (1995).
}
\label{fig2}
\end{minipage}
\end{figure}

\section{Formation of High-Mass Black Holes}

We find that the black holes in transient sources can be formed
from stars with ZAMS masses in the interval $20-35\msun$
(Brown, Lee, \& Bethe 1999).
The black hole mass is only slightly smaller than the He core mass,
typically $\sim 7\msun$ (Bethe, Brown, \& Lee 1999).

Crucial to our discussion here is the fact that single stars evolve very
differently from stars in binaries that lose their H-envelope
either on
the main sequence (Case A) or in the giant phase (Case B). However, stars that
transfer mass or lose mass after core He burning (Case C) evolve, for our
purposes, as single stars, because the He core is then exposed too close to
its death for wind mass loss to significantly alter its fate.
The core masses of single stars and binary stars are summarized in
Fig.~\ref{fig1}. 
Single stars above a ZAMS mass of about $20\msun$ skip convective
carbon burning following core He burning, with the result, as we shall
explain, that their Fe cores
are substantially more massive than stars in binaries, in which
H-envelope has been transferred or lifted off before He core burning.
These latter ``naked" He stars burn $^{12}C$ convectively,
and end up with relatively
small Fe cores. The reason that they do this has to do chiefly with the
large mass loss rates of the ``naked" He cores, which behave like W.-R.'s.
In the ZAMS mass range $\sim 20-35\msun$,
it is clear that many, if not most, of the single stars go into
high-mass black holes, whereas stars in binaries which burn ``naked" He
cores go into low-mass compact objects. In this region of
ZAMS masses the use of high He-star mass loss rates does not cause
large effects (Wellstein \& Langer 1999).

The convective carbon burning phase (when it occurs) is
extremely important in pre-supernova evolution,
because this is the first phase in which a large amount of entropy
can be carried off in $\nu\bar\nu$-pair emission, especially if
this phase is of long duration.  The reaction in which carbon burns is
$^{12}C(\alpha,\gamma)^{16}O$ (other reactions like $^{12}C+^{12}C$ 
would require
excessive temperatures).  The cross section of $^{12}C(\alpha,\gamma)^{16}O$ is
still not accurately determined; the lower this cross section the higher the
temperature of the $^{12}C$ burning, and therefore the more intense the
$\nu\bar\nu$ emission.
With the relatively low
$^{12}C(\alpha,\gamma)^{16}O$ rates determined both directly from nuclear
reactions and from nucleosynthesis by Weaver \& Woosley (1993), the
entropy carried off during $^{12}C$ burning in the stars of ZAMS
mass $\le 20\msun$ is substantial.
The result is rather low-mass Fe cores for these stars, which can evolve
into neutron stars. Note that in the literature earlier than
Weaver \& Woosley (1993) often large $^{12}C(\alpha,\gamma)^{16}O$
rates were used, so that the $^{12}C$ was converted into oxygen and the
convective burning did not have time to be effective.
Thus its role was not widely appreciated.

Of particular importance is the ZAMS mass at which the convective carbon
burning is skipped.
In Fig.~\ref{fig2}, this occurs at
ZAMS mass $19\msun$ but with a slightly lower $^{12}C(\alpha,\gamma)^{16}O$
rate it might come at $20\msun$ or higher. As the progenitor
mass increases, it follows from general polytropic arguments that the
entropy at a given burning stage increases.
At the higher entropies of the more massive stars the density at which
burning occurs is lower, because the temperature is almost fixed for a
given fuel. Lower densities decrease the rate of the triple-$\alpha$
process which produces $^{12}C$ relative to the two-body
$^{12}C(\alpha,\gamma)^{16}O$ which produces oxygen.
Therefore, at the higher entropies in the more massive stars the
ratio of $^{12}C$ to $^{16}O$ at the end of He burning is lower.
The star skips the long convective carbon
burning and goes on to the much shorter oxygen burning.  Oxygen burning goes
via $^{16}O + ^{16}O$ giving various products, at very much higher temperature
than $C(\alpha,\gamma)$ and much faster.
Since neutrino cooling during the long carbon-burning phase gets
rid of a lot of entropy of the core, skipping this phase leaves
the core entropy higher and the final Chandrasekhar core fatter.
We believe that our above discussion indicates that single stars in the region
of ZAMS masses $\sim 20-35\msun$ end up as high mass black holes.

Arguments have been given that SN 1987A
with progenitor ZAMS mass of $\sim 18\msun$ evolved into a low-mass
black hole (Brown \& Bethe 1994). We believe from our above arguments
that just above the ZAMS mass of $\sim 20\msun$, single
stars go into high-mass black holes without return of matter to the
Galaxy.
Thus, the region of masses for low-mass black hole formation in
single stars is narrow, say $\sim 18 -20\msun$ (although we believe
it to be much larger in binaries).

\section{Quiet Black Hole - Main Sequence Star Binaries}

We believe that there are many main sequence stars more massive than the
$\lsim 1\msun$ we used in our schematic evolution, which end up further
away from the black hole and will fill their Roche Lobe during only
subgiant or giant stage. From our evolution, we see that a $2\msun$ main
sequence star will end up about twice as far from the black hole as
the $1\msun$, a $3\msun$ star, three times as far, etc. Two of the
9 systems in our Table \ref{tab1} have subgiant donors
(V404 Cyg and XN Sco). These have the longest periods, 6.5 and 2.6 days
and XN Sco is suggested to have a relatively massive donor of $\sim 2\msun$.
It seems clear that these donors sat inside their Roche Lobes until they
evolved off the main sequence,
and then poured matter onto the black hole once they expanded and
filled their Roche Lobe. For a $2\msun$ star, the evolutionary time is
about a percent of the main-sequence time, so the fact that we see two subgiants
out of nine transient sources means that many more of these massive donors
are sitting quietly well within their Roche Lobes.
Indeed, we could estimate from the relative time, that there are
$2/9\times 100 =22$ times more of these latter quiet main sequence stars
in binaries.

\section{Discussion}

We have shown that it is likely that single stars in the range of ZAMS
masses $\sim 20-35\msun$ evolve into high-mass black holes without return of
matter to the Galaxy. This results because at mass $\sim 20\msun$ the
convective carbon burning is skipped and this leads to substantially
more massive Fe cores. Even with more realistic reduced mass loss
rates on He stars, however,
it is unlikely that stars in this mass range in binaries
evolve into high-mass black holes,
because the progenitor of the compact object when stripped of its
hydrogen envelope in either Case A (during main sequence)
 or Case B (RLOF) mass transfer will burn as a
``naked" He star, ending up as an Fe core which is not sufficiently
massive to form a high-mass black hole.

In the region of ZAMS mass $\sim 40 \msun$, depending sensitively on
the rate of He-star wind loss, the fate of the single star or
the primary in a binary may
be a low-mass black hole.
In our estimates we have assumed the Brown \& Bethe (1994) estimates of
$1.5\msun$ for maximum neutron star mass and $1.5-2.5\msun$ for the range
in which low-mass black holes can result.

In our evolution of the transient sources using Case C (during He shell 
burning) mass transfer, almost the entire He core will collapse into a 
high-mass black hole (Bethe, Brown, \& Lee 1999), explaining
the more or less common black hole mass of $\sim 7\msun$ for these objects,
with the possible exception of V404 Cygni where the mass may be greater.
Our evolution gives an explanation for the seemingly large gap in
black-hole masses, between the $\gsim 1.5\msun$ for the black hole we believe
was formed in 1987A and the $\sim 1.8\msun$ black hole we suggest in
1700-37 and the $\sim 7\msun$ in the transient sources.

We note that following the removal of the H envelope by Case C mass
transfer, the collapse inwards of the He envelope into the developing
black hole offers the Collapsar scenario for the most energetic
gamma ray bursters of MacFadyen \& Woosley (1999).

\acknowledgements
We would like to thank Charles Bailyn and Stan Woosley
for useful discussions.
We were supported by the U.S. Department of Energy under Grant No.
DE--FG02--88ER40388.

\end{document}